\documentclass[preprint,5p]{elsarticle}

\usepackage[colorlinks=true,linkcolor=blue]{hyperref}
\usepackage{tikz}
\usepackage{amsmath}
\usepackage{graphicx}
\usepackage{array}
\usepackage{multirow}
\usepackage{subfigure}
\usepackage{xcolor}
\usepackage{balance}
\usepackage{amsfonts}
\usepackage{amssymb}
\usepackage{algorithm}
\usepackage{algorithmic}
\usetikzlibrary{shadows,trees}
\usepackage{longtable}
\tikzstyle{every node}=[draw=black,anchor=west,rounded corners=0.05cm,font=\footnotesize,fill=white,drop shadow]
\journal{Journal of \LaTeX\ Templates}

\begin{document}
\begin{sloppypar}
\begin{frontmatter}

\title{Is Blockchain for Internet of Medical Things a Panacea for COVID-19 Pandemic?}

\author[label1]{Xuran Li}
\author[label2]{Bishenghui Tao}
\author[label2]{Hong-Ning Dai \corref{cor1}}
\ead{hndai@ieee.org}
\author[label3]{Muhammad Imran}
\author[label4]{Dehuan Wan}
\author[label1]{Dengwang Li }

 \address[label1]{Shandong Key Laboratory of Medical Physics and Image Processing, School of Physics and Electronics, Shandong Normal University, Jinan, Shandong, China; sdnulxr@sdnu.edu.cn; dengwang@sdnu.edu.cn}

 \address[label2]{Faculty of Information Technology, Macau University of Science and Technology, Macau SAR;

 2009853YII30001@student.must.edu.mo, hndai@ieee.org}
 \address[label3]{College of Applied Computer Science, King Saud University, Riyadh, Saudi Arabia; dr.m.imran@ieee.org}
 \address[label4]{Guangdong University of Finance, Guangzhou, China; wan\underline{~~}e@gduf.edu.cn}
\cortext[cor1]{Corresponding author}

\begin{abstract}
The outbreak of the COVID-19 pandemic has deeply influenced the lifestyle of the general public and the healthcare system of the society. As a promising approach to address the emerging challenges caused by the epidemic of infectious diseases like COVID-19, Internet of Medical Things (IoMT) deployed in hospitals, clinics, and healthcare centers can save the diagnosis time and improve the efficiency of medical resources though privacy and security concerns of IoMT stall the wide adoption.  In order to tackle the privacy, security, and interoperability issues of IoMT, we propose a framework of blockchain-enabled IoMT by introducing blockchain to incumbent IoMT systems. In this paper, we review the benefits of this architecture and illustrate the opportunities brought by blockchain-enabled IoMT. We also provide use cases of blockchain-enabled IoMT on fighting against the COVID-19 pandemic, including the prevention of infectious diseases, location sharing and contact tracing, and the supply chain of injectable medicines. We also outline future work in this area.
\end{abstract}

\begin{keyword}
Blockchain, Internet of Medical Things, Security, Privacy, COVID-19
\end{keyword}

\end{frontmatter}


\section{Introduction}
\label{sec:intro}
Since the year 2020, the whole world has been struggling to fight against the spread of the new coronavirus disease called SARS-CoV-2 (aka, COVID-19). COVID-19 is a dangerous respiratory infection that rapidly spreads from humans to humans. Even though many countries are sparing their efforts to slow down or control the outbreaks of COVID-19 and gradually restore regular lives, we still have a long way to go before the COVID-19 pandemic being fully solved. The widespread of the COVID-19 pandemic has led to an adverse influence on almost all aspects of human life and has also exposed the limitations on medical sources of the current healthcare systems.

Since the emergence of the COVID-19 pandemic, substantial efforts have been paid on exploiting technological advances to effectively fight against this disease. Among all these technologies, the Internet of Medical Things (IoMT) is one of the most promising approaches to help to prevent the widespread of this infectious disease~\cite{MOHDAMAN:JNCA2020,hndai:IOTJM20}. IoMT is a remote healthcare system, which is mainly composed of medical sensor devices, medical data servers, and professional medical staff (i.e., doctors, nurses, and so on). The medical sensor devices (including various biomedical sensors, RFID tags, and QR tags) collect the medical data of patients and then transmit the data to medical data servers. The authorized professional medical staff can then access the medical data, conduct an early diagnosis, and provide medical therapy measures. Meanwhile, IoMT can monitor the status of patients so as to provide professional medical support.

The wide adoption of IoMT systems to incumbent medical institutions and agencies brings potentials to address the COVID-19 pandemic. First, the time of waiting for an early diagnosis can be saved. The early diagnosis for COVID-19 is necessary since the contagiousness rate of COVID-19 is very high~\cite{Habibzadeh:IIoT2020}. However, the medical resources (such as hospitals and doctors) are limited and the chances of early diagnosis for many patients are missed. When physiological data of a potential patient is collected by medical sensor devices, the professional medical staff can soon diagnose if the potential patient is infected. Second, monitoring patients with medical sensors can reduce the possibility of medical staff being infected. The possibility of direct contact between patients and medical staff can be minimized with IoMT-enabled remote healthcare service. Last but not least, limited medical resources can be saved. When a patient gets out of emergency status and only needs more rest for recovery, the medical staff can monitor this patient at given time intervals with the help of IoMT. The constrained medical resources (i.e., hospital beds, medical devices, and medical staff) can be saved for those patients in emergency status.

Though the adoption of IoMT is beneficial to combat COVID-19, there are some challenges to be solved before IoMT can be widely deployed. Due to the limited computing capability and battery capacity, complicated encryption algorithms are not feasible for most wearable or implanted medical sensor devices. Consequently, wireless data transmission of IoMT can be vulnerable to external attacks~\cite{YSun2019:IA,WRafique:ICST2020}. Since medical data contains sensitive personal information of patients, it is crucial to allow the authorized users to access the medical data while preserving the privacy of patients~\cite{MShen:MNET19}. In addition, there are diverse types of medical sensor devices~\cite{xrLi:ComCom2020} and heterogeneous IoMT networks, thereby leading to the poor interoperability of IoMT.

One promising solution to the aforementioned challenges is integrating blockchain with IoMT~\cite{Ray:ISJ2020,NGarg:IA2020,hndai:IOTJM20}. The inherent nature of blockchain includes decentralization, trustworthiness, traceability, and transparency. Therefore, blockchain can potentially address the security, privacy, and interoperability challenges~\cite{MOIN:FGCS2019,JWan:ITII2019}. Firstly, the decentralized blockchain can protect the sensitive medical data of patients from being fully controlled by third-party entities. Secondly, the decentralization of blockchain can also avoid the single point of failure and mitigate the bottleneck at central servers due to the increasing number of medical sensor devices. Thirdly, blockchain-based IoMT can guarantee the security and traceability of IoMT data because the contents on the blockchain will not be controlled by any single entity, and the medical data as well as event logs stored on the blockchain are immutable~\cite{Ali:IComST2019}. Fourthly, the decentralized peer-to-peer (P2P) network architecture can help to process the heterogeneous IoMT data and improve the interoperability of IoMT.

The convergence of blockchain and IoMT may improve the security of IoMT, enhanced the privacy protection of IoMT data, and ameliorate the interoperability of IoMT systems. Hence, we explore the convergence of blockchain and IoMT in this paper. In particular, we propose a framework for integrating blockchain with IoMT, investigate the solutions of blockchain-enabled IoMT to tackle the COVID-19 pandemic, discuss the potential applications of blockchain-enabled IoMT, and outline future directions in blockchain-enabled IoMT. The main contributions of this paper are summarized as follows:
\begin{itemize}
\item We present the technical overview of blockchain and IoMT. Specifically, we introduce the blockchain structure, consensus algorithms, smart contract, and categories of blockchain systems.
\item We propose a framework of integrating blockchain with IoMT and analyze the potentials brought by this framework. In particular, blockchain helps to improve the security of IoMT, enhance privacy protection of IoMT, and ameliorate the interoperability of IoMT systems.
\item We provide use cases of blockchain-enabled IoMT in combating the COVID-19 pandemic, including the prevention of infectious diseases, location sharing and contact tracing, and the supply chain of injectable medicines.
\end{itemize}

The rest of this paper is organized as follows. Section~\ref{sec:related} provides related studies in literature. Section~\ref{sec:overview} presents the technological review of blockchain and IoMT. Section~\ref{sec:BIoMT} introduces an architecture of blockchain-enabled IoMT. The case studies of blockchain-enabled IoMT are provided in Section~\ref{sec:app_BIoMT}. The future directions are summarized in Section~\ref{sec:future}. Section~\ref{sec:conclu} concludes this paper.

\begin{figure}[t]
\scalebox{0.88}{
\begin{tikzpicture}[%
  grow via three points={one child at (0.4,-0.7) and
  two children at (0.4,-0.7) and (0.4,-1.4)},
  edge from parent path={(\tikzparentnode.south)|-(\tikzchildnode.west)}]
  \node {This paper}
    child { node {Section 1: \hyperref[sec:intro]{Introduction}}}		
    child { node {Section 2: \hyperref[sec:related]{Related Work}}}
    child { node {Section 3: \hyperref[sec:overview]{Overview of Blockchain and IoMT}}
      child { node {2.1: \hyperref[]{Blockchain}}}
      child { node {2.2: \hyperref[]{Internet of Medical Things}}}
    }
    child [missing] {}				
    child [missing] {}	
    child { node {Section 4: \hyperref[sec:BIoMT]{Blockchain-enabled IoMT}}
      child { node {4.1: \hyperref[]{Architecture of BCeMT}}}
      child { node {4.2: \hyperref[]{Opportunities of BCeMT}}}
      child { node {4.3: \hyperref[]{Challenges of BCeMT}}}
    }
    child [missing] {}				
    child [missing] {}
    child [missing] {}
    child { node {Section 5: \hyperref[sec:app_BIoMT]{Applications of BCeMT}}
      child { node {5.1: \hyperref[ ]{Location sharing and contact tracing}}}
      child { node {5.2: \hyperref[ ]{Prevention of infectious diseases}}}
      child { node {5.3: \hyperref[ ]{Traceable information chain of vaccines}}}
    }
    child [missing] {}				
    child [missing] {}
    child [missing] {}				
    child { node {Section 6: \hyperref[sec:future]{Future directions}}
      child { node {6.1: \hyperref[ ]{IoMT data sharing}}}
      child { node {6.2: \hyperref[ ]{Scalability and optimization}}}
      child { node {6.3: \hyperref[ ]{Big data analytics of BCeMT}}}
      child { node {6.4: \hyperref[ ]{Integration of AI with BCeMT}}}
      }
    child [missing] {}				
    child [missing] {}
    child [missing] {}
    child [missing] {}
    child { node {Section 7: \hyperref[sec:conclu]{Conclusion}}};
\end{tikzpicture}
}
\caption{Structural diagram of the paper.}
\label{fig:structure}
\end{figure}
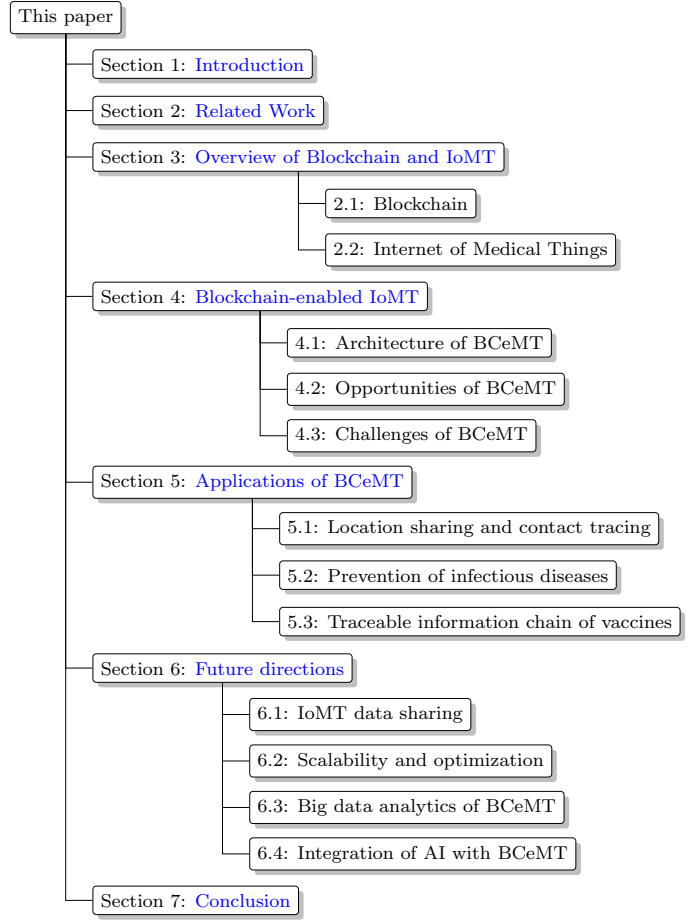

\section{Related Works}
\label{sec:related}

\begin{table*}[ht]
\caption{The comparison of related works with this work}
\label{tab:compare_related}
\scriptsize
\renewcommand{\arraystretch}{1.5}
\centering
\begin{tabular}{|c|p{6.5cm}|p{1.5cm}|p{2cm}|p{2cm}|p{1.7cm}|}\hline
{\bf References}&{\bf Main contributions}&{\bf Blockchain}& {\bf IoMT}&{\bf Integrating blockchain with IoMT}&{\bf Use cases in combating COVID-19}\\
\hline\hline
\cite{Azana2021:JNCA}  &  Presenting the architecture of the IoMT medical ecosystem and the applications of IoMT systems in combating COVID-19 &  $\times$  &   $\checkmark$    &    $\times$ & $\checkmark$   \\
\hline
\cite{Yang2020:Diag}  & Proposing a point-of-care diagnostics based IoMT platform for COVID-19 diagnosis and monitoring &  $\times$  &   $\checkmark$    &    $\times$ & $\checkmark$ \\
\hline
\cite{Ravi2020:JCOT}  & Proposing the IoMT system for mobility restricted orthopedic patients during COVID-19 pandemic &  $\times$  &   $\checkmark$    &    $\times$ & $\checkmark$ \\
\hline
\cite{Klaine2020:ITM}  & Proposing a privacy-preserving framework for contact tracing based on the blockchain &   $\checkmark$  &   $\times$    &   $\times$   & $\checkmark$ \\
\hline
\cite{Kumar2021:TJS}  & Designing and implemented the security and privacy IoMT framework based on the blockchain and interplanetary file system &  $\checkmark$  &   $\checkmark$    &   $\checkmark$   & $\times$ \\
\hline
\cite{Fakhri2020:SCS}  & Presenting a system architecture of integrated blockchain with IoMT and developed an app based on this architecture &  $\checkmark$  &   $\checkmark$    &   $\checkmark$   & $\times$ \\
\hline
\cite{Ray2021:ISJ}  & Proposing a data-flow architecture for the integration of blockchain and IoMT and presenting the use cases &  $\checkmark$ &   $\checkmark$    &   $\checkmark$   & $\times$ \\
\hline
\cite{RenY2019:Sen}  & Proposing a blockchain-enabled storage mechanism in the cloud-assisted WBAN &  $\checkmark$  &   $\checkmark$    &   $\checkmark$   & $\times$ \\
\hline
\cite{Son2020:IA}  & Designing the authentication protocol for the system integrating cloud-assisted WBAN with blockchain &  $\checkmark$ &   $\checkmark$    &   $\checkmark$   & $\times$ \\
\hline
\cite{Uddin2019:WiMob}  & Proposing a blockchain-based lightweight consensus mechanism for a cloud-assisted WBAN enabled remote patient monitoring system & $\checkmark$ &   $\checkmark$    &   $\checkmark$   & $\times$ \\
\hline
This paper  & Proposing a framework of integrating blockchain with IoMT and analyze the potentials brought by this framework, presenting use cases in combating the COVID-19 pandemic & $\checkmark$ &   $\checkmark$    &   $\checkmark$   &$\checkmark$ \\
\hline
\end{tabular}
\end{table*}

The outbreak of the COVID-19 pandemic has brought severe influence on the whole world due to the high infectiousness of COVID-19. One of the potential solutions to address the COVID-19 pandemic is deploying IoMT across medical institutions as well as communities. The authors in ~\cite{Azana2021:JNCA} present a comparison between the traditional medical ecosystem and IoMT medical ecosystem and discuss the applications of IoMT systems in combating COVID-19 in different countries. In another work~\cite{Yang2020:Diag}, a point-of-care diagnostics-based IoMT platform is proposed for patients infected by COVID-19. With this platform, the patients can dynamically monitor the disease status themselves and receiving medical support without spreading the virus. Meanwhile, the IoMT system for providing medical support to orthopedic patients with limited mobility during this COVID-19 pandemic environment is designed in~\cite{Ravi2020:JCOT}. However, most medical sensor devices in IoMT systems are limited in terms of computing capability and battery capacity. Moreover, there are heterogeneous types of medical sensor devices, leading to the complexity of the IoMT system. The main challenges of deploying the IoMT lie in 1) difficulty in assuring the security of patient's data and 2) the poor interoperability between medical sensor devices.

One promising solution for addressing these challenges of the IoMT system is blockchain. The research~\cite{Klaine2020:ITM} presents a blockchain-enabled framework for digital contact-tracing applications in the COVID-19 pandemic. In this framework, both the trust and privacy of users are guaranteed by the blockchain. Another research investigates enhancing the security and privacy of medical systems with blockchain in~\cite{Kumar2021:TJS}. In this research, the blockchain and interplanetary file systems are utilized for constructing the security and privacy framework of IoMT. The authors in~\cite{Fakhri2020:SCS} investigate the performance improvement of the blockchain network and present the system architecture of utilizing blockchain in IoMT. They develop a smartphone app for the automation of medical records based on their proposed architecture. The integration of blockchain and IoMT is also investigated in~\cite{Ray2021:ISJ}. In this research, a data-flow architecture is proposed for the integration of blockchain and IoMT, and the corresponding use cases are presented.

Considering the limitation of the storage capability of blockchain and the massive amount of physiological data collected from patients, the integration of cloud-assisted wireless body area network (WBAN) with blockchain in the smart medical system is investigated in studies~\cite{RenY2019:Sen, Son2020:IA,Uddin2019:WiMob}.
In~\cite{RenY2019:Sen}, a blockchain-enabled storage mechanism in the cloud-assisted WBAN is proposed. Meanwhile, the authors in~\cite{Son2020:IA} investigate a telecare medical information system, which is implemented in the cloud-assisted WBAN. To guarantee the security of patient health data, they also design the blockchain-based authentication protocol for this system. The work~\cite{Uddin2019:WiMob} investigates a cloud-assisted WBAN enabled remote patient monitoring system. In this system, they designed a lightweight consensus mechanism based on blockchain.

Different from the aforementioned researches, this paper proposes the framework of integrating blockchain with IoMT and analyzes the potential brought by this framework, especially in the context of COVID-19. In addition, we provide use cases of this framework in combating the COVID-19 pandemic, including the prevention of infectious diseases, location sharing and contact tracing, and the supply chain of injectable medicines. The comparison of the related studies with this paper is shown in Table~\ref{tab:compare_related}.

\section{Overview of Blockchain and IoMT}
\label{sec:overview}

This section presents a technical overview on both blockchain and IoMT.
\subsection{Blockchain}

In 2009, Nakamoto put forward Bitcoin, this brand new digital currency without any authoritative intermediary's coordination. In the system of Bitcoin, people who do not trust each other can directly make deal with this cryptocurrency~\cite{nakamoto2019bitcoin}. In December 2013, a new blockchain-framework platform Ethereum was proposed by Buterin~\cite{wood2014ethereum}. In addition to the built-in algorithms based on digital currency transactions, Ethereum also provides the Turing-complete programming language for the smart contract, which was the first applied to the blockchain.

\begin{figure}[t]
\centering
\includegraphics[width=8.8cm]{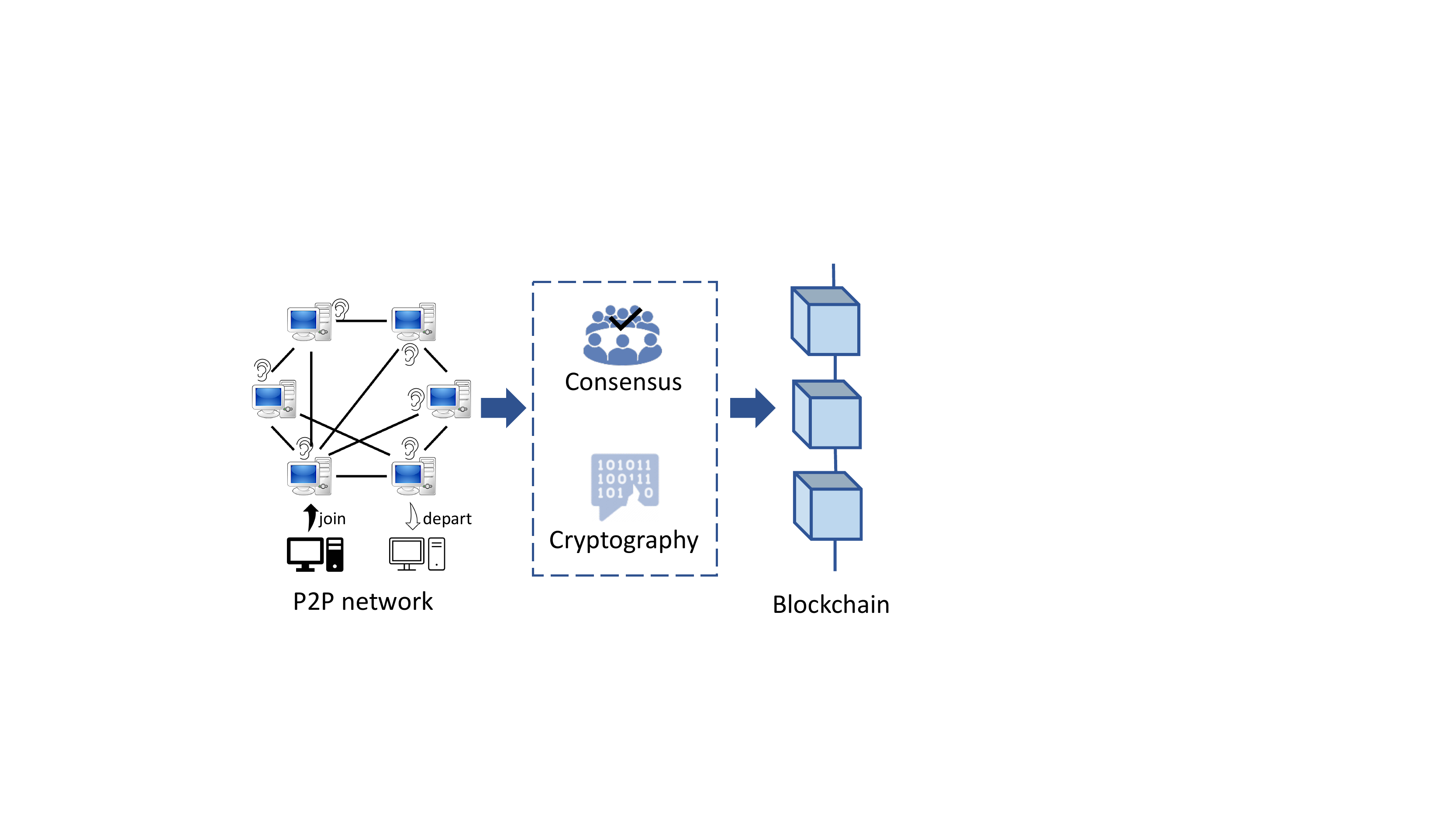}
\caption{Blockchain technology system.}
\label{fig:Btch}
\end{figure}

The blockchain has a distinctive data structure. Information is encapsulated into every single block, and the blocks are sequentially combined into a chain. Block cipher is used to ensure that the data cannot be altered and forged. The system is decentralized, implying that there is no central node in the network of blockchain. Thus, any two nodes can trade with each other directly. Hence, blockchain networks mostly choose P2P protocol as the network transmission protocol. In this way, the blockchain system can be totally distributed so that the failures of a single point are tolerated.

The nodes in the blockchain network generally have key characteristics of equality, autonomy, distribution, and so on~\cite{zheng2018blockchain,Hakak:IN2020}. Moreover, as shown in Figure~\ref{fig:Btch}, every node can freely join or depart from this system. During the information propagation, every single node always listens to the broadcasting messages in the network. Once a node receives a new transaction or block from neighbor peers, the broadcast message will be verified. The message includes a digital signature, proof of work, hash value, and so on. In order to prevent the invalid data from recording, only when the verification of the information has been to a consensus, the message would be encrypted and appended into the blockchain.

\subsubsection{Blockchain structure}

\begin{figure}[t]
\centering
\includegraphics[width=8.8cm]{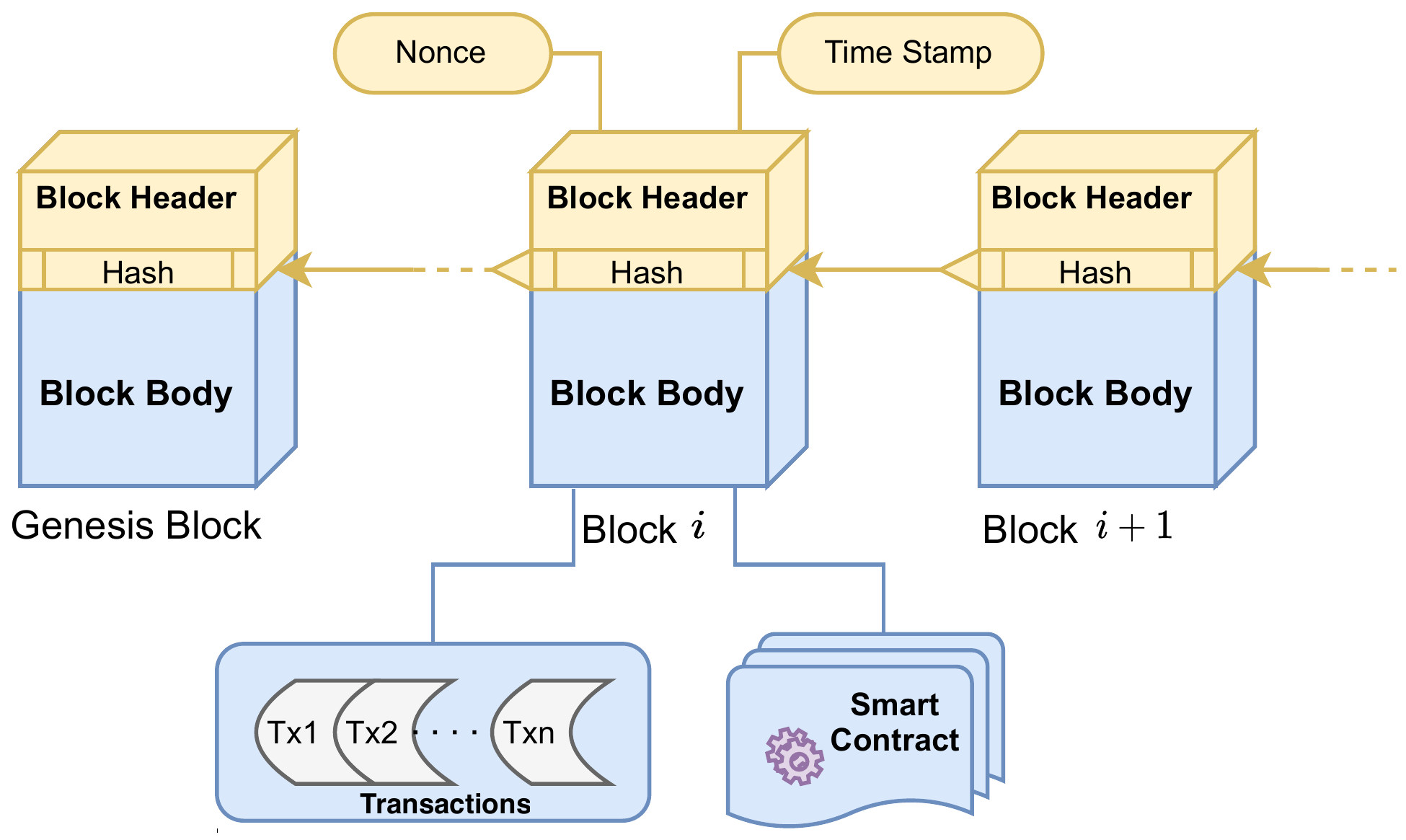}
\caption{Blockchain structure.}
\label{fig:struc}
\end{figure}

The structure of a blockchain is built as a connected block list. In this chain-like list, every single block has a hash pointing to the previous one. A correct (or confirmed) chain contains a complete block list and the transactions. It is available for every node to download and maintain the chain, implying that the blockchain system acts like a distributed public ledger. Figure.~\ref{fig:struc} presents an instance of the blockchain structure. Each block in a chain consists of a block body and a block header, with a hash pointer contained in the header pointing to its prior block, which is also called its parent block. And the initial block which does not have any parent block is also called genesis block~\cite{zheng2018blockchain}. Some information for identification and marking will be stored in the block header, such as Nonce and Time Stamp. Meanwhile, the majority of complete data is stored in the block body, such as transaction data, smart contracts, etc.

A block contains a block header and a block body. Generally, as listed in Table~\ref{tab:block}, a block header includes a block version, Merkle tree root hash, timestamp, $n$Bits, nonce, and the previous block hash~\cite{zheng2017overview}. It is worth mentioning that the information in the block cannot be modified by others with a hash pointer in the block header. The block body contains validated transactions as well as a transaction counter. Take the Bitcoin block as an example. The maximum size of a single block is defined as 1 MB and a transaction usually has a size of around 250 bytes. Thus, each block has a limit of 4,000 transactions. According to the real-time data of the Blockchain Explorer, the recent number of transactions of Bitcoin is about 2,000 transactions per block~\cite{koops2018predicting}. Moreover, the average Ethereum block size is between 20 to 30 KB mostly~\cite{rouhani2017performance}.

\begin{table}
\caption{An example of block header components and description.}
\renewcommand{\arraystretch}{1.45}
\centering
\resizebox{0.4\textwidth}{!}
{
\begin{tabular}{|m{3cm}|m{5cm}|}
\hline
\textbf{Components}                & \textbf{Description}                                                       \\ \hline\hline
Block   Version & Indicates   which version of protocol and validation rules that the block defers to.   \\ \hline
Merkle Tree Root Hash & The   hash value of every transaction within the block.          \\ \hline
Time Stamp            & Current   time as seconds in universal time since January 1, 1970.  \\ \hline
$n$Bits                 & Current   hashing target in a compact format.                       \\ \hline
Nonce         &    random number, usually starts with 0 and increases for every hash calculation.\\ \hline
Previous block hash   & A   256-bit hash value of the previous block.                      \\ \hline
\end{tabular}
}
\\
\label{tab:block}
\end{table}

\subsubsection{Consensus algorithms}
Various consensus algorithms are deployed to determine how to achieve agreement when verifying and recording new transactions and blocks. An efficient consensus mechanism needs to allow all participants to reach an agreement in a non-trusted environment and maintain the system under a good fault tolerance. There are different consensus algorithms in different platforms. As the first and the widest deployed consensus mechanism, the Proof of Work (PoW) introduced by the Bitcoin network assumes that all candidates take part in racing with their computing power, to find a required nonce value to construct the right block~\cite{nakamoto2019bitcoin}. PoW guarantees a decentralized network and a public ledger system for trustless entities dealing with each other. However, PoW suffers from excessive power consumption. 

Proof of Stake (PoS) can potentially overcome the drawbacks of PoW. The PoS comes from the concept that the more stake coins a node has, the higher chance it can fabricate the new block~\cite{nguyen2018survey}. Therefore, nodes with more tokens are generally believed to be more inclined to maintain the security of the network in order to protect their own rights. PoS avoids a massive consumption of computing power, thereby being more energy-efficient than PoW. However, PoS still confronts the risk of attacks due to its low mining cost.

Delegated proof of stake (DPoS) is an extension of the PoS. This consensus protocol supports the users who have stakes to vote the delegate or witnesses, to build the blocks and chain, or change the parameters of the network. DPoS demonstrates better operational efficiency and double-spending attack protection. However, DPoS can result in a centralized network due to a limited number of selected delegates.

To protect the system from the attacks of potential malicious nodes, Hyperledger Fabric~\cite{androulaki2018hyperledger} utilizes the Practical byzantine fault tolerance(PBFT)~\cite{castro1999practical}. In PBFT, replications among nodes reach consensus despite the failing or incorrect information propagation in the distributed network so as to enable the Byzantine faults tolerance. Moreover, after introducing the idea of voting nodes to record the transactions, delegated byzantine fault tolerance (DBFT) is implemented for saving communication consumption. In addition, Ripple protocol requires less trust to maintain the consensus with low latency, by utilizing collectively-trusted subnetworks from the larger network~\cite{schwartz2014ripple}, thereby showing robustness when facing the Byzantine failures.

\subsubsection{Smart contract}

A smart contract is a protocol or a computing program deployed on the blockchain to automatically execute, control, or verify actions under the agreement between different parties. The conception of the smart contract was first proposed by Szabo in the 1996~\cite{szabo1996smart} to reduce the requirement of trust, cost of enforcement, and exceptions of malicious attacks during the transaction. On top of the blockchain system, every user can call and interact with smart contracts to conduct various business activities, such as making transactions with others, receiving and sending messages, and voting activities~\cite{ZZheng:FGCS2020}. Smart contracts share similar characteristics to the blockchain, such as distribution, decentralization, and immutability. Once a smart contract is deployed, no one can modify it, ensuring the security of transactions and systems. For an instance, Ethereum implements smart contracts in various computer languages like Solidity, Serpent, or LLL~\cite{wang2018overview}. When a smart contract is called, it will run immediately in the content of an Ethereum Virtual Machine (EVM) on the decentralized network computers.

\subsubsection{Categories of Blockchain systems}

The current blockchain systems can be categorized as public blockchains, private blockchains, and consortium blockchains. The access permission and network properties vary from different blockchain systems. we next briefly describe three typical blockchain systems.

\textit{Public Blockchains}: Every user within a public blockchain system can freely take part in and interact with the blockchain. Blockchain-based cryptocurrency systems such as Bitcoin~\cite{nakamoto2019bitcoin}, Ethereum~\cite{wood2014ethereum}, and Litecoin have this mechanism. With no central node in charge of this decentralized and trustless network, the cryptography algorithm and consensus ensure the validity and integrity of data, thereby maintaining the fairness and equality of the system.

\textit{Private Blockchains}: Unlike public blockchains, not anyone has access permission in private blockchain systems. Only authorized entities can access the on-chain data and initiate transactions, whereas the blockchain is essentially a centralized database. Private blockchains have generally been used for data confidentiality, authentication, and organization management of internal information.

\textit{Consortium Blockchains}: As a combination of public blockchains and private blockchains, consortium blockchains have typically been under the control of some reign entities. The access mechanism is determined by manager nodes to decide who can join the network, initiate transactions, or participate in the consensus. Hyperledger Fabric~\cite{androulaki2018hyperledger}, Quorum~\cite{chase2016quorum}, and Corda~\cite{brown2016corda} are the typical examples of consortium blockchains.

\subsubsection{Summary}
Blockchain technologies have various irreplaceable characteristics that are quite suitable for the application of the IoMT. For example, the immutability and traceability of blockchains can bring innovations to the data storage, security, and privacy protection of the IoMT. It makes the tracing and tracking of medical incidents more efficient while also avoiding privacy leakage and tampering attacks. Furthermore, nodes in the blockchain system can join and exit the system freely; this feature is very suitable for flexible connections of perception or actuation devices. Moreover, the application of smart contracts can provide new solutions for medical services and data management in IoMT,  such as epidemic spread control, vaccination statistics, and medical equipment management.

Besides, among the various categories of blockchain systems, they have corresponding appropriate application scenarios. Due to the sensitivity of physiological data from patients, the private blockchain that a single group or multiple groups control (e.g., doctors, patients, administrators) is the most feasible type of blockchains for the IoMT applications. Besides, compared with public blockchains, private blockchains have better performance in scalability and flexibility. Patients' physiological data are not of high sensitivity for some scenarios, and the consortium blockchains may also be feasible.

In short, the integration of blockchain technology with the existing medical information system is a promising direction. Undeniably, it will also make new contributions to combat against COVID-19.

\subsection{Internet of Medical Things}
\label{subsec:IoMT}
The system architecture and existing challenges of IoMT are presented in the following subsections.

\begin{figure}[t]
\centering
\includegraphics[width=8.8cm]{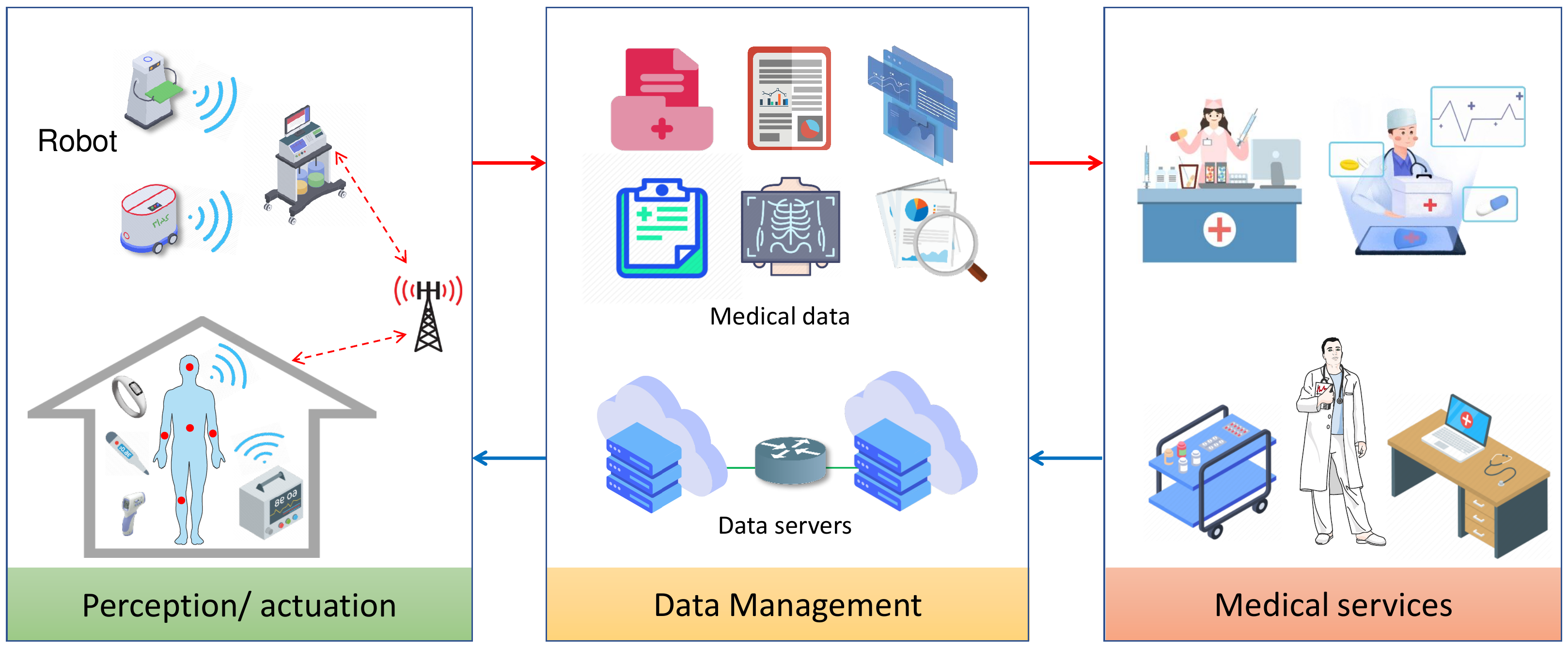}
\caption{Overview of IoMT.}
\label{fig:IoMT}
\end{figure}

\subsubsection{System architecture of IoMT}

The recently proposed IoMT-based medical system architecture usually consists of three layers~\cite{xrLi:ComCom2020,Qadri:CST20}. As shown in Figure~\ref{fig:IoMT}, the three layers of IoMT architecture are composed of the data collection layer, the data management layer and the medical service layer.

\textbf{Perception/Actuation layer:}
On the one hand, the perception/actuation layer collects the physiological data of patients and preprocess these physiological data with patients' privacy information. On the other hand, this layer can also enforce some actions on the IoMT. This layer is mainly composed of two types of devices: medical sensor/actuator devices and edge servers. Medical sensing devices are wearable or implanted biosensors, which are used to collect physiological data of patients, such as blood sugar, blood pressure, heart rate, electrocardiogram (ECG), electromyography (EMG) and electroencephalogram (EEG), etc. The medical sensor devices conduct the on-demand monitoring and collect the patient's physiological data at set intervals when the patients' physiological parameters are normal. When the patient's physiological parameters become abnormal or under the request of an authorized user (patient, doctor, nurse, etc.), the medical sensor devices will conduct continuous intensive monitoring of the patient.

After collecting the patients' physiological data, the medical sensor devices transmit the collected patient's physiological data to the edge servers. Edge servers may be gateways in the network, local base stations, or personal smart devices. The main function of edge servers is to preprocess the collected physiological data with various formats, eliminate the redundancy of the data, compress and encrypt these data. The edge servers can also conduct preliminary analysis and storage of the data, and generate alarming signals when physiological parameters become abnormal. When the edge servers complete the preliminary processing of the collected patients' physiological data, the data is uploaded to the data management layer through WiFi, LTE or 5G network.

\textbf{Data management layer:}
In the data management layer, the patients' physiological data uploaded by edge servers are further processed, stored and analyzed on the data management equipment. The data management equipment in this layer needs to effectively manage and analyze the heterogeneous physiological data of patients, and classify the data according to the timeliness and the priority of the analysis task. In addition, the data management equipment needs to establish an efficient distributed storage mechanism for a large amount of data to support efficient data processing and analysis. In the scenario of intensive care, sudden disease detection and vital parameters monitoring, the real-time changes of physiological parameters such as heart rate and blood pressure reflect the health status of individuals. Therefore, these data need to be processed in a very short time, and then the analytical results on patients' physiological data should be rapidly returned to the medical professionals and the patient to deal with the emergencies.

Since the data management equipment needs to analyze patients' physiological data efficiently and rapidly, some recent studies employ cloud servers to conduct data processing, data storage and data analysis in this layer~\cite{JAbawajy2017:ICMag,RCao2020:IIoT}. The cloud servers extract features of the collected physiological data and classify the data to help medical professionals (such as doctors) providing better medical service. In this way, medical professionals can obtain the processed data with the easy observed form, and provide prediction as well as treatment advice faster and better. With the permission of the patient, the algorithms and programs for the assessment of expected diagnosis and rehabilitation progress can be run on the data management equipment. In this layer, access control mechanisms or identity authentication mechanisms are applied to ensure that only authorized entities can access patient's physiological data, so that data security and patients' privacy are protected. In addition, patients' information and related profiles will be anonymous before sharing data with other entities such as research centers.

\textbf{Medical service layer:}
The medical service layer aims to provide users with basic visual data analysis results. The visualized data are used to generate reports and send them to healthcare participants (including doctors, patients, and nursing staff) involved in clinical observation, patient diagnosis, and intervention processes. Medical personnel (such as doctors) with appropriate authentication and authorization credentials can access the reports on patient's physiological data and provide medical advice to patients timely.

The medical professionals can track designated patients and access the reports based on the daily activities of patients. On the one hand, through real-time analysis, the changes in patients' physiological parameters can be detected immediately to avoid sudden diseases. When the patient's physiological parameters become abnormal, the medical service equipment sends a notification alarm to the nursing staff for further treatment advice. Then the medical professionals could guard against risk and initiate the necessary action plan to deal with emergencies such as heart attack, falls, etc. On the other hand, through long-term monitoring and analysis of patient physiological data, the medical professionals can track the health status of patients' daily life and predict some potential health risks, such as obesity and hypertension. In addition, considering the physiological characteristics of individuals, individual treatment simulations can be carried out to assess the health risk and design the best medical plan.

\subsubsection{Challenges in IoMT}

The advent of IoMT still confronts the following challenges.

\textit{Absence of interoperability:}
There are a variety of medical sensor devices in the data collection layer of IoMT, where the medical sensor devices are different in computing capability, memory, energy supply and embedded software. The data formats of collected patient's physiological data vary from device to device, consequently resulting in the difficulty of managing the data in the data management layer.

Another reason caused the poor interoperability lies in the heterogeneity of wireless/wired protocols. When the patient's physiological data is collected, the medical sensor devices transmit the collected data to edge servers. During this transmission process, different wireless protocols are employed. For some powerful wearable medical sensor devices, the protocols like WiFi, NB-IoT and Bluetooth Low Energy (BLE)  are utilized. The protocols like Near-field communication (NFC), Radio Frequency Identification (RFID) are employed for medical sensor devices implanted into the patient's body.  The poor interoperability across different IoMT systems will result in the difficulty of medical information exchange in different medical institutions.

\textbf{Privacy and security:}
The leakage of patients' sensitive physiological information may lead to serious problems to patients. However, most medical sensor devices in the data collection layer are resource-constrained. The limitations of computing power, memory, energy supply in medical sensor devices also result in the infeasibility of traditional complicated encryption mechanisms. The patient's sensitive physiological data from medical sensor devices to edge serves may be wiretapped by illegal devices.

To protect the privacy of patients' sensitive information in the data management layer and health service layer, various access control mechanisms and identity authentication mechanisms are designed~\cite{YSun2019:IA}. However, the data storage of patients' sensitive physiological data still relies on the third party, where the bugs on information leakage may exist in the data server. Consequently, data storage or cloud services providers may mistakenly or intentionally release patients' privacy-sensitive data.

\section{Blockchain-enabled IoMT}
\label{sec:BIoMT}

In this section, we propose an architecture for incorporating blockchain into IoMT systems, namely BCeMT, and investigate the benefits as well as challenges of BCeMT.

\subsection{Architecture of Blockchain-enabled IoMT}
\begin{figure}[t]
\centering
\includegraphics[width=8.8cm]{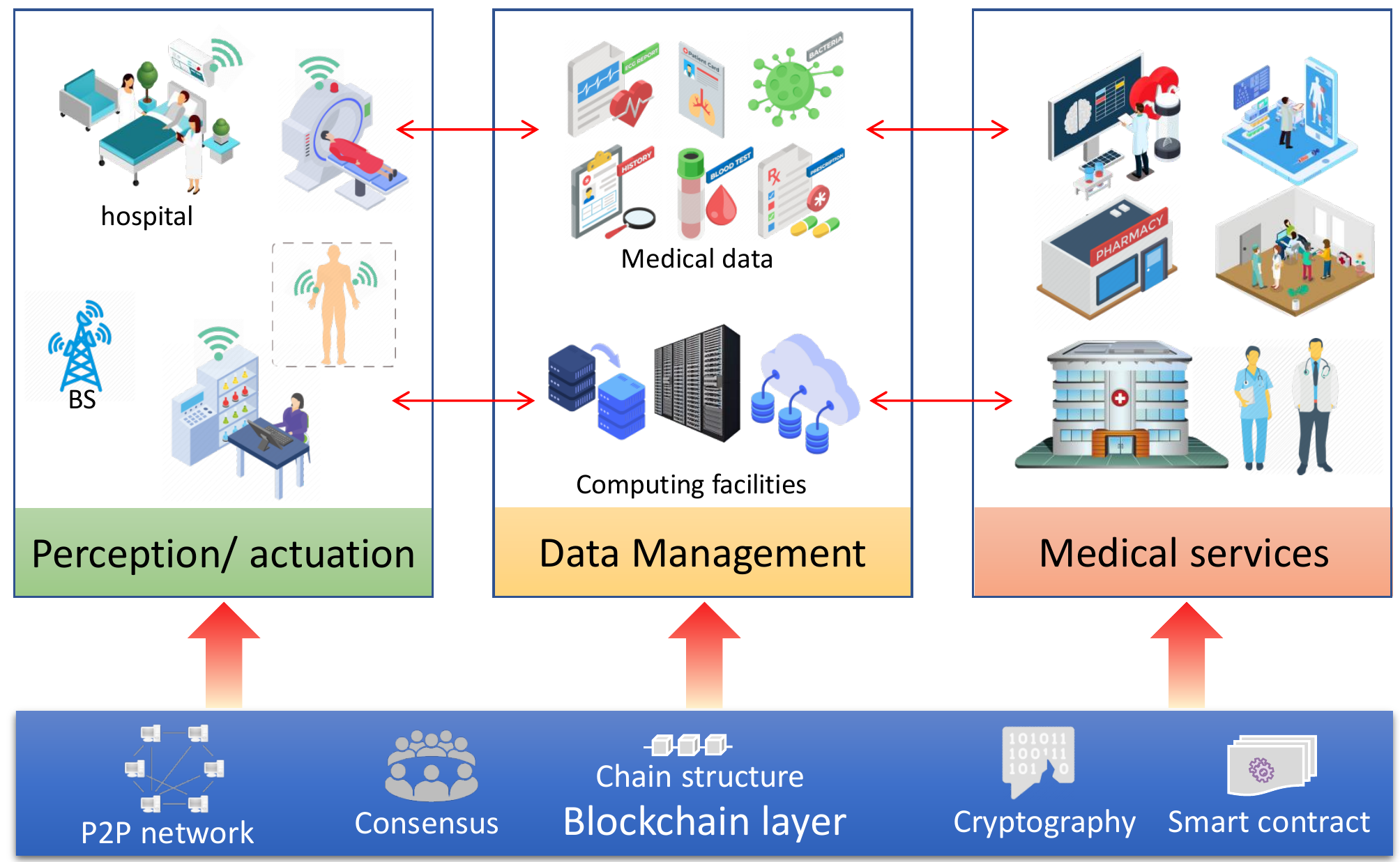}
\caption{System architecture of BCeMT.}
\label{fig:biomt}
\end{figure}

The convergence of blockchain and IoMT can ameliorate the interoperability of IoMT systems, significantly improve the security of IoMT, and enhance the privacy protection of IoMT~\cite{hndai:IOTJM20}. Figure~\ref{fig:biomt} depicts a system architecture of BCeMT. In this architecture, the blockchain layer is essentially serving for the entire IoMT, i.e., covering the perception/actuation layer, data management, and medical services layer.

In this architecture, the blockchain is serving as a crucial infrastructure connecting different layers of IoMT. Blockchains are endowed with cryptographic schemes (i.e., digital signature and public encryption), P2P networks, distributed consensus, smart contracts, and a chain of blocks. Consequently, the blockchain can ensure a certain security of IoMT. The employment of the authentication~\cite{YWu:MNet20}, homomorphic obfuscations, and group signature~\cite{SZhang:IOTJ20} to blockchains can further protect the data privacy of IoMT. Moreover, the overlaid P2P networks in blockchain systems can connect different sectors in IoMT together so as to improve the interoperability across the entire IoMT.

\subsection{Opportunities of BCeMT}

BCeMT can overcome challenges of IoMT and offer the following opportunities to enhance IoMT.

\subsubsection{Interoperability improvement}

The introduction of blockchain to IoMT can essentially connect fragmented sectors of IoMT together via the overlaid P2P network, which is a crucial part of blockchain systems. Thereafter, different IoMT sectors, such as medical centers, clinics, hospitals, and homes can be connected together to offer a ubiquitous Internet service across the entire IoMT~\cite{YWu:IoTJ20}. Therefore, the interoperability across different healthcare sectors has been improved.

In addition, this novel architecture can also help to collect, preprocess, and store diverse IoMT data. During this process, heterogeneous IoMT data can be converted, transformed, encrypted, compressed, and stored in blockchain (i.e., on-chain data)~\cite{dai-iotj-2019}; alternatively IoMT data can be saved in an off-chain fashion, in which only hashes of IoMT data are stored in blockchain~\cite{eberhardt2017or}. Consequently, the interoperability of IoMT has also been greatly improved.

\begin{table*}[t]
\centering
\caption{Applying blockchain technique for privacy preserving of IoMT}
\label{tab:BP}
\scriptsize
\renewcommand{\arraystretch}{1.5}
\begin{tabular}{|c|p{2cm}|p{3.5cm}|p{3.2cm}|p{5.5cm}|}\hline
{\bf References}&{\bf Blockchain type}&{\bf Design objectives}& {\bf Main contributions}&{\bf Methods}\\
\hline\hline
\cite{GDagher:SCS2018}  & permissioned &  protecting the privacy of patients' EHR data    &   access control scheme with proxy re-encryption method   & storing keys and encrypted records on the blockchain \\
\hline
\cite{Pussewalage:IThing2018} & not mentioned & addressing the privacy issues of patients' EHR data  & attribute based access control scheme &   managing the attribute assignments and delegation control with blockchain  \\
\hline
\cite{SSaha:ICC2020} &  private & preserving the privacy of patients &  access control mechanism  &  ECC-enabled signature scheme\\
\hline
\cite{VRamani:GLOBECOM2018}& private or permissioned  & medical data accessibility is controlled by the patient  & data accessibility mechanism & lightweight public key cryptographic operations \\
\hline
\cite{AOmar:FGCS2019}  & permissioned   & protect the privacy of patients' EHR data & a platform to store the sensitive personal information on blockchain &  ECC cryptographic function is applied to achieve the pseudonymity of patients \\
\hline
\cite{RGuo:IA2018}  &  not mentioned & ensuring the anonymity of patients and immutability of data in EHRs  & attribute-based signature mechanism & introducing multiple authorities to this system and sharing pseudorandom function seed in every two authorities
and keep secretly\\
 \hline
\cite{QSu:IA2020}  & permissioned  & privacy preservation both patients and doctors  &  attribute-based signature scheme & utilizing the KUNodes algorithm for revocation \\
\hline
\cite{JXu:IIoT2019} & public  & protecting both patients' medical data and the doctors' diagnoses &  a two blockchains enabled privacy preserving framework & one blockchain used for sharing data and the other used for fine-grained access control \\
\hline
\cite{ZTong:JMS2019}  & permissioned and permission-less  & protecting the privacy of patients during the medical data sharing process &  a framework consists of two blockchains & a permissioned blockchain stores the medical data and patiens' personal information, and send the medical data to the permission-less blockchain at set intervals \\
\hline
\cite{XLiang:PIMRC2017}  & permissioned  & allowing patients to selectively share the medical data & An App for BCeMT & the access control list of medical data is generated according to the settings and operations of patients\\
\hline
\cite{XYue:JMS2016}   & private & ensuring patients to manage their own medical data by themselves & An App for BCeMT & integrating the traditional database with gateway, and blockchain helps to store medical data \\
\hline
\end{tabular}
\end{table*}

\subsubsection{Privacy preservation of IoMT data}

To preserve the privacy of patients, only authorized users are allowed to access the sensitive medical data of patients.
The research~\cite{Pussewalage:IThing2018} proposed an attribute-based access control scheme to address the privacy issues of patients' Electronic health record (EHR) data in the blockchain-based e-health system. In this scheme, the attribute assignments and delegation control are managed by the blockchain in a lightweight style. Meanwhile, the research~\cite{GDagher:SCS2018} proposed a permissioned blockchain-enabled framework and an access control scheme to protect the privacy of patients' EHR data. In their framework, the distributed proxy re-encryption method and the corresponding re-encryption contract are utilized. With the proxy re-encryption method, keys as well as small encrypted records can be stored on the blockchain.

Due to the limited storage capacity of IoMT, the lightweight cryptographic operations in BCeMT are investigated. Particularly, the authors of ~\cite{SSaha:ICC2020} designed a private blockchain-based access control mechanism for the applications of IoMT. The privacy preservation of this mechanism is achieved by the elliptic curve cryptography (ECC) enabled signature approach. Moreover, in~\cite{VRamani:GLOBECOM2018}, a blockchain-based data access mechanism is proposed to guarantee that medical data accessibility is controlled by the patients. In this mechanism, the lightweight public key cryptographic operations are conducted with ECC cryptographic function. The work~\cite{AOmar:FGCS2019} proposed a platform to protect the privacy of patients' EHR data by storing the sensitive information on the blockchain. In this platform, the ECC cryptographic function is applied to achieve the pseudonymity of patients.

In addition to access control schemes, attribute-based signature mechanisms can be utilized to preserve the privacy of patients. In particular, in~\cite{RGuo:IA2018}, the authors designed an attribute-based signature mechanism for a blockchain-based EHR system. To ensure the anonymity of patients and immutability of data in EHRs, this mechanism introduced multiple authorities. Another attribute-based signature scheme in a blockchain-based EHR system is proposed in ~\cite{QSu:IA2020}. While in this mechanism, the attributes are revocable by utilizing the KUNodes algorithm, and therefore the identity privacy of users (both patients and doctors) is protected.

Utilizing more than one blockchain is another possible solution to deal with the privacy leakage problem of IoMT. The research~\cite{JXu:IIoT2019} proposed a privacy-preserving framewrork enabled by two blockchains for the smart healthcare system. In this framework, one blockchain is used for publishing data and the other is used for fine-grained access control, while the medical data is stored in a file system. In this way, neither the patients' medical data nor the doctors' diagnoses will be tampered. The authors in~\cite{ZTong:JMS2019}  proposed another two-blockchains enabled privacy-preserving framework. In this framework, a permissioned blockchain is applied to store the medical data as well as patients' personal information, and periodically anchors the medical data to a permission-less blockchain. Medical data sharing is enabled and the privacy of patients is protected.

Some researches focus on designing Applications (Apps) to make sure that patients could manage their own medical data by themselves. In~\cite{XLiang:PIMRC2017}, the authors designed a mobile App in a permissioned blockchain-enabled medical data sharing system. With this App, the patients shall selectively share the medical data according to the necessity of their own judgment. The research ~\cite{XYue:JMS2016} proposed another App based on an architecture that integrating the traditional database with the gateway. In this architecture, the blockchain helps to store medical data and patients are clearly aware of the usage of medical data.

A summary of the researches on utilizing blockchain technique to preserve the privacy of patients in IoMT is given in Table \ref{tab:BP}.

\begin{table*}[t]
\centering
\caption{Applying blockchain technique for security assurance of IoMT}
\label{tab:BS}
\scriptsize
\renewcommand{\arraystretch}{1.5}
\begin{tabular}{|c|p{2cm}|p{3.7cm}|p{3.2cm}|p{5.5cm}|}\hline
{\bf References}&{\bf Blockchain type}&{\bf Design objective}& {\bf Main contributions}&{\bf Methods}\\
\hline\hline
\cite{NGarg:IA2020} & public &protecting the security of patients' medical data in BCeMT & authenticated key management protocol & one-way cryptographic hash function and bitwise XOR operations \\
\hline
\cite{AYazdinejad:JBHI2020}  & public  & protecting the security of patients' medical data in BCeMT &  lightweight decentralized authentication scheme  & lightweight symmetric key encryption algorithm ARX \\
\hline
\cite{ZWang:JPDC2020} & consortium & ensure the security of medical data &  a BCeMT framework  & two smart contracts are utilized to securely share and store the medical data individually \\
 \hline
\cite{HLPham:GCW2018} & not mentioned &  protect patients' personal and information and medical data  & a remote medical system based on blockchain  & securing medical data with smart contract \\
 \hline
\cite{MShen:IN2019}   & not mentioned  & protect the patients' sensitive information security in the medical image retrieval process &  a BCeMT system for medical image retrieval process & storing the extracted and encrypted feature vectors of each image into the blockchain \\
\hline
\cite{VPatel:HIJ2019} & not mentioned  & ensuring the patients to control their identity information by themselves in medical images sharing process &  a BCeMT framework for medical image retrieval process   &  identities of patients are stored on the blockchain as randomly-generated public keys\\
\hline
\cite{PZhang:CSBJ2018}& public   & securing the medical data sharing system & a blockchain enabled medical framework   & utilizing the patients' digital medical identities to encrypt the to be shared medical data\\
 \hline
\end{tabular}
\end{table*}

\subsubsection{Security assurance of IoMT}
To protect the sensitive medical data of IoMT from being wiretapped, the medical data is required to be encrypted before the transmission and retrieval process.
In~\cite{NGarg:IA2020}, the authors designed an authenticated key management protocol for a BCeMT environment with lightweight cryptographic operations. Specifically, the one-way cryptographic hash function and bitwise XOR operations are applied to conduct the lightweight cryptographic operations for computing resources limited medical sensor devices. Their protocol helps to protect the security of patient's medical data, and the security performance of this protocol is verified by the tool AVISPA. The work~\cite{AYazdinejad:JBHI2020} proposed a lightweight decentralized authentication scheme for patients in distributed IoMT with a public blockchain. In this authentication scheme, the symmetric key encryption algorithm ARX is utilized to conduct lightweight encryption operations on the data of medical sensor devices.

The smart contract can be utilized to help secure the blockchain-based IoMT. The authors in~\cite{ZWang:JPDC2020} proposed a blockchain-enabled framework to ensure the security of medical data. In this framework, two smart contracts are applied to securely share and store the medical data individually, and a graph neural network based trust model is used to detect the malicious nodes. The proxy re-encryption method is utilized in their scheme so that patients shall dynamically provide or revoke the permissions of their medical data access. Meanwhile, in~\cite{HLPham:GCW2018}, the authors introduced the process of securing medical data with smart contract in a blockchain-enabled remote healthcare system.

In some cases, the volume of medical data is too large to be stored in the blockchain, so that only the vital part of the data will be stored on the blockchain.
In~\cite{MShen:IN2019}, the authors proposed a BCeMT system to protect the patients' sensitive information security in the medical image retrieval process. The feature vectors of each image are extracted and encrypted before storing them in the blockchain. The integrity and accuracy of encrypted image features will be verified by miners with a hash signature when the information is uploading to the blockchain. Another research~\cite{VPatel:HIJ2019} also utilizes blockchain technology to protect the patients' sensitive information in medical images. In this research, the identities of patients are stored on the blockchain as randomly generated public keys, and the patients' identity information is controlled by themselves in medical image sharing.

In~\cite{PZhang:CSBJ2018}, the authors proposed a blockchain-enabled framework for securing the medical data sharing system. In this framework, the patients' digital medical identities are utilized to encrypt the medical content to be shared and the security of medical data is enhanced. Based on this framework, a case study of decentralized App for sharing medical data is demonstrated.

A summary of the researches on utilizing blockchain techniques to protect the security of patients' data in IoMT is given in Table \ref{tab:BS}.

\subsection{Challenges of BCeMT}
Although the BCeMT architecture brings a number opportunities, there are still a few challenges that remain to be addressed before BCeMT can be practically adopted.

\subsubsection{Resource Constraints}

Storage and computation capability constraints may limit the widespread applications of BCeMT architecture. On one hand, a massive amount of physiological data may be generated from the IoMT network, thereby posing stringent requirements on high storage and computation capability. Moreover, the chain in the blockchain network has kept growing and each node stores a replica of the entire chain. On the other hand, the processing ability of each node in the blockchain network is also limited~\cite{dai-iotj-2019}.

One potential solution to this challenge is to introduce cloud services into this architecture. In this way, we can upload a large amount of less privacy-sensitive data to the remote clouds while storing the sensitive data (e.g., the sensitive private information of patients) at the blockchain. Processing most medical data on the clouds will also release the computation burden at nodes in the blockchain.

\subsubsection{Governance issues}

The current emerged private and permissioned blockchain applications are granted ``write'' permissions to a predefined peer or set of peers~\cite{Reyna2018:FGCS}. In the sensitive healthcare domain, the legal regulations on the BCeMT are required to explicitly define the nodes, users, peers and validators of the blockchain.

For governance, only participating hospitals and institutions, as well as selected patients who are permissioned, are included in the blockchain network~\cite{Mackey2019:BMC}. The adoption of BCeMT architecture needs to be backed by legal instruments.


\section{Applications of BCeMT}
\label{sec:app_BIoMT}

BCeMT can potentially address the crisis caused by the COVID-19 epidemic. We list several possible applications of BCeMT in tackling the COVID-19 crisis.

\subsection{Location sharing and contact tracing}

\begin{figure}[h]
\centering
\includegraphics[width=8.8cm]{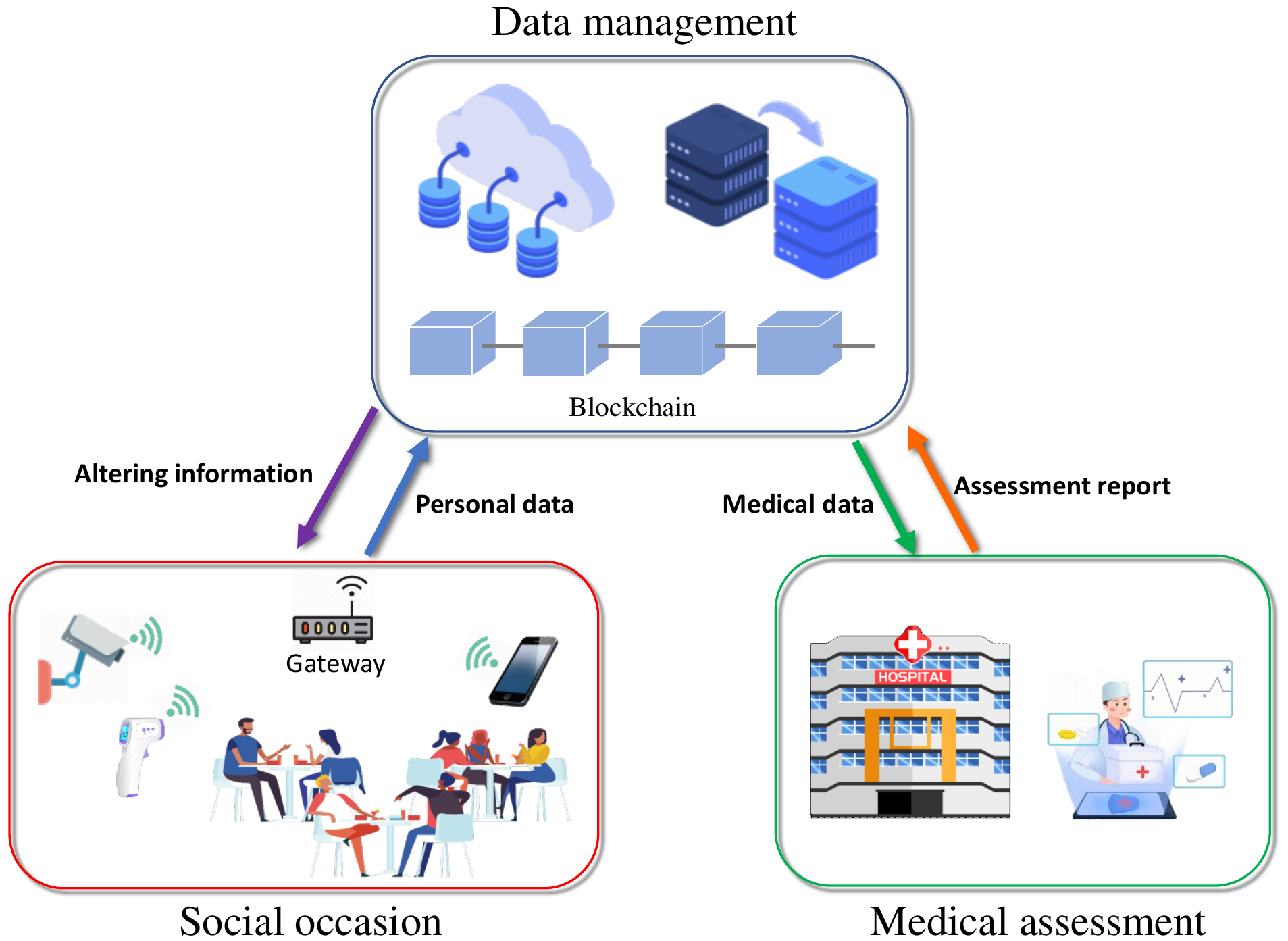}
\caption{Location sharing and contact tracing.}
\label{fig:BIoMTCT}
\end{figure}

According to recent studies ~\cite{ALHURAIMEL:STTE2020,KUMAR:STTE2020}, there are two main transmission pathways of the COVID-19 virus. The first is the direct human-to-human transmission through respiratory droplets from the coughing, sneezing or even breathing of patients. The second is an indirect pathway through touching the staff surface infected by patients. Therefore, obtaining the location information of patients and tracing their close contacts are vital to prevent the transmission of the COVID-19 virus. However, the conventional information-sharing methods may lead to privacy leakage and the patients may lose control of their personal data.

The blockchain-based IoMT may solve this problem with the decentralized architecture and unforgeability characteristic. As shown in Figure~\ref{fig:BIoMTCT}, the potential patients and general public in the social occasion first upload the location and medical information to the blockchain-based database, where the patients' information pseudonymity can be achieved by cryptographic functions~\cite{AOmar:FGCS2019}. Next, the authenticated medical centers will obtain the medical data and conduct the COVID-19 assessment. If the assessment result is positive, the blockchain-based database will connect to the medical center for medical support and inform the gateway on social occasions to generate the alarming signal to warn the general public. Privacy preservation can be achieved with the location sharing scheme~\cite{YJi:JMS2018} and the transparency feature of blockchain may guarantee the unforgeability of contact traceability.

\subsection{Prevention of infectious diseases}

\begin{figure}[t]
\centering
\includegraphics[width=8.8cm]{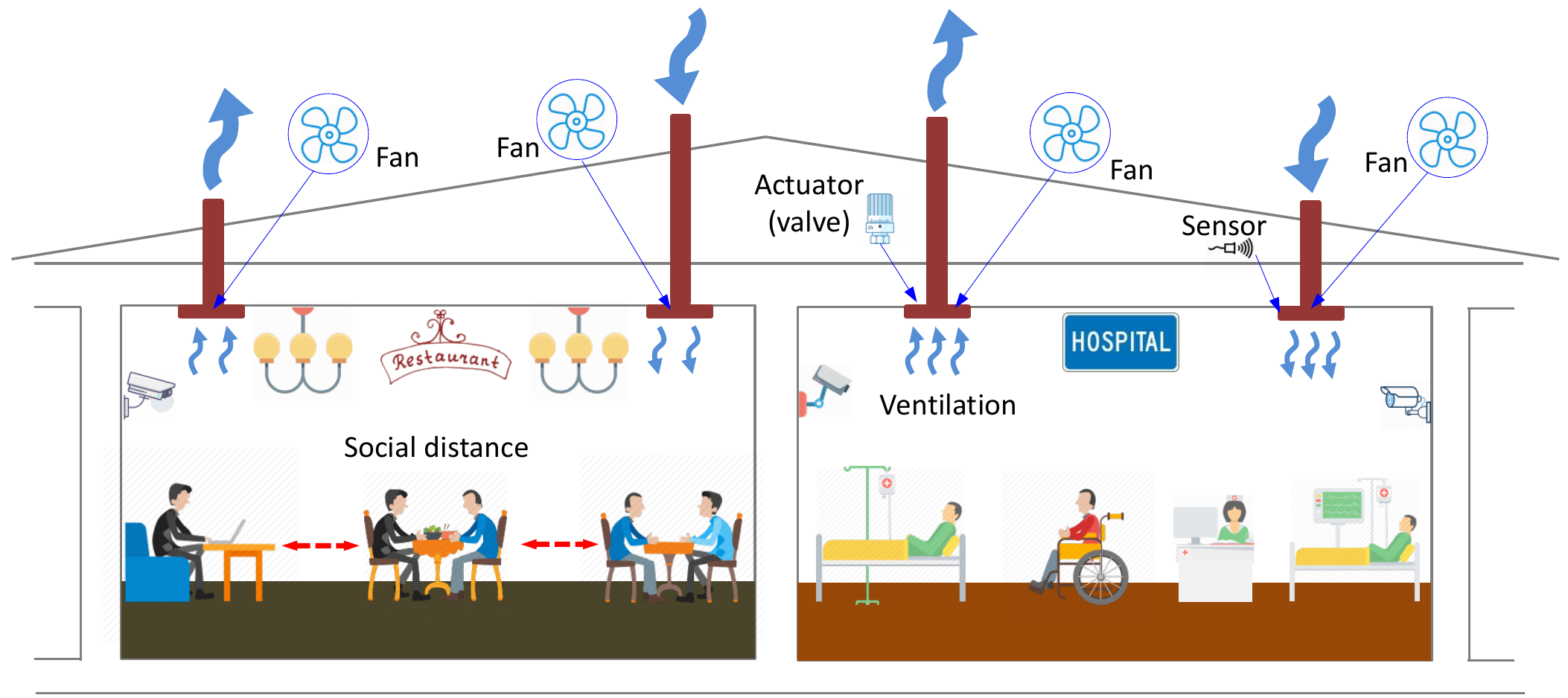}
\caption{Prevention of infectious diseases.}
\label{fig:prevent}
\end{figure}

As indicated in \cite{lewnard2020scientific}, keeping a social distance is an effective way to prevent the outbreaks of infectious diseases like COVID-19. BCeMT can offer a solution to social-distancing measures. For example, as shown Figure~\ref{fig:prevent}, sensors as well as cameras can detect, identify, and count the number of customers in a restaurant, consequently offering early warnings of an overcrowded environment. Moreover, \cite{lu2020covid} shows that the increment of air ventilation in a crowded environment can also diminish the aerosol transmission of coronavirus. The controller of a ventilation fan or an air conditioner can dynamically adjust the ventilation volume according to the crowd density which can be obtained by cameras and ambience sensors. The ventilation fans or central air conditioning systems can be shut down or reduced airflow for the sparse crowd so as to save energy consumption.

The outbreaks of COVID-19 also result in overcrowded hospitals. It becomes extremely important to improve air filtration inwards or emergency rooms in a hospital. In addition, it is also crucial to ensure the functioning of exhaust fans in restroom facilities of a hospital so as to reduce the air transmission risk of coronavirus. However, it is laborious for technicians to troubleshoot every fan and air conditioner in the entire ventilation system. In addition, it can also cause health risks for technicians to manually check the ventilation system due to the contaminated surfaces and filters~\cite{MOUCHTOURI:2020113599}. In this case, the wide proliferation of diverse sensors and actuators (like controllers of fans) in IoMT can offer a solution to this emerging issue. On the one hand, sensors can report the possible faults of fans and abnormal functioning of the ventilation system. On the other hand, actuators of fans can dynamically adjust airflow volumes as shown in Figure~\ref{fig:prevent}.

\subsection{Traceable information chain of COVID-19 vaccines}

Undoubtedly, vaccines play an indispensable role in the control of the COVID-19 epidemic. Researchers around the world have invested enormous effort in vaccine development, testing, and mass production as soon as possible to fight against the spread of the virus. Due to the huge demand for this vaccine, we can reasonably speculate that there will be many problems in the mass production and distribution of injections, such as quality control, transportation, and storage safety. In fact, some suspected adverse reactions after injection have also been found in existing tests and injection results~\cite{KAUR2020198114}. Therefore, it is necessary to design and build a vaccine supply chain system based on the blockchain to ensure the quality, safety, and traceability of the vaccine.  When a problem occurs, this vaccine-tracing system can quickly trace back and locate the source of the problem. In addition, due to massive COVID-19 vaccine production, the production of other vaccines may also be affected due to excessive production lines, raw materials, and manpower requirements. The system combined with IoMT can also provide data support for the rational distribution of productivity.

\begin{figure}[t]
\centering
\includegraphics[width=8.2cm]{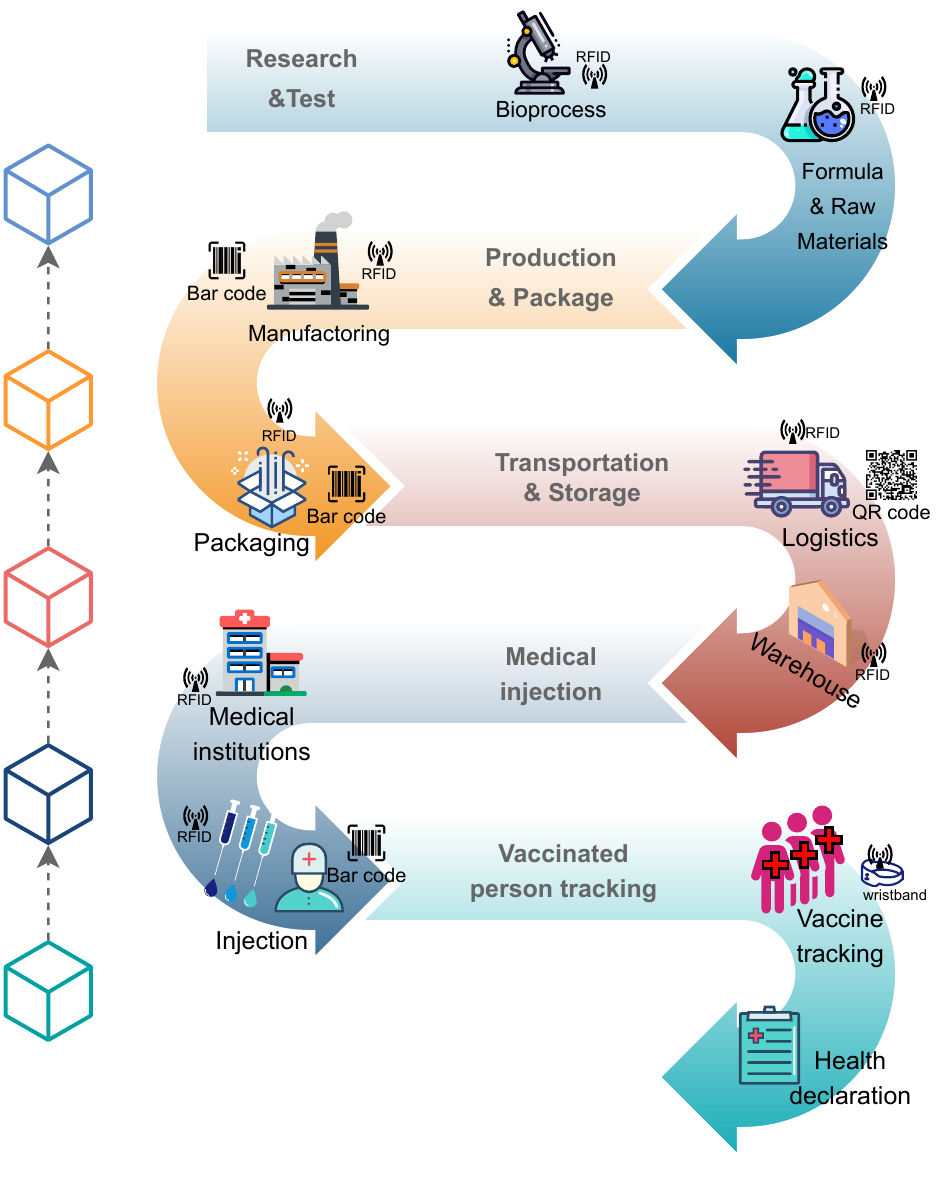}
\caption{Supply chain of injectable medicines.}
\label{fig: Supply_chain}
\end{figure}

The immutability and security of the blockchain are suitable for data recording and storage in the injectable medical supply chain. As shown in Figure~\ref{fig: Supply_chain}, from vaccine research and testing to follow-up observation of vaccine injections, and health declaration, information can be collected and recorded on the chain. In the operation and maintenance, various IoMT devices can evolve, such as sensors, Radio Frequency Identification (RFID) tags, bracelets, etc~\cite{hndai:IOTJM20}. In the IoMT blockchain system as illustrated in Figure~\ref{fig: Supply_chain}, vaccine information at different stages will be encapsulated on the chain, including the results of research and testing, the source of production materials, process factories, transportation logistics and storage warehouse information, hospitals and medical staffs, the follow-up symptom observation after the injection. Among them, RFID tags, Quick Response tags (QR tags), Bar code tags, and sensors can ensure that information recording is more efficient and non-falsifiable. The patients after vaccine injections can declare their health status regularly during the observation period, thereby ensuring the subsequent collection of injections more complete.


\section{Future directions}
\label{sec:future}

Although BCeMT has the great potential to address the COVID-19 crisis, there are a number of issues to be tackled. We discuss the open issues as follows.

\subsection{Legislation and incentives of IoMT data sharing}

Although BCeMT is promising in fostering interoperability of IoMT systems, offering privacy preservation of IoMT data, and guaranteeing the security of IoMT systems, incumbent medical institutions, organizations and the public have their concerns or misgivings to share IoMT data. The first reason is the absence of regulations and the legislation of medical data~\cite{milne2019trust}. Secondly, another obstacle in medical data sharing lies in patients' concerns on how their medical data is shared and exploited~\cite{Jihoon:2019.9550}.

To dispel the public's misgivings of IoMT data sharing, there are several working directions in the future. Firstly, substantial legislative efforts are necessary to be made to regulate IoMT data sharing and data governance. Clear definitions and regulations on how to share IoMT data, which part of IoMT data to be shared, and which party to use IoMT data. During this process, blockchain may also serve a crucial role in promoting the regulations and standardization of IoMT data sharing. For example, traceable blockchain can make the data-sharing process be fully traceable so as to improve the transparency of data sharing and governance. Secondly, blockchain can play as a catalyst for data sharing. For example, built-in incentive/pricing mechanisms of blockchain can be adopted to encourage the public or patients to share their medical data. Medical and research agencies can pay for the shared IoMT data. It is worth mentioning that data privacy protection mechanisms are still necessary for medical data sharing.

\subsection{Scalability and optimization of blockchain}

The emerging BCeMT also poses stringent requirements on blockchains in terms of throughput and storage. Thus, the scalability, throughput efficiency, and modularity of blockchains will become valuable research directions in the future.

Both the scalability and throughput efficiency of the blockchain have received extensive attention~\cite{YWu:IoTJ20}. The massive IoMT data may overload the existing blockchain systems. There are several possible solutions to this issue.  1) \emph{Adopting new blockchain data storage method}, for example, the combination of on-chain and off-chain data storage can avoid the problem of huge on-chain data redundancy and difficulty to synchronize. 2) \emph{Efficiency of consensus algorithms}, we can adjust existing consensus algorithms after finding the best performance consensus parameters, such as encryption algorithm, block size, block interval, etc. 3) \emph{hybrid blockchain type}, a flexible consortium chain integrating the private chain or public chain may possibly increase the transaction volume per unit time.

Moreover, the modularity of blockchains can also offer a flexible manner to disassembling and reconstructing blockchain systems so as to support diverse applications (especially for COVID-19) without affecting the security of blockchain data. The modularity can be achieved by designing suitable smart contract deployment plans.


\subsection{Privacy-preserving big data analytics of BCeMT}
In the IoMT system, a large amount of medical data will be continuously collected by the medical sensor devices. With the help of big data technology, the collected medical data in IoMT may be utilized more effectively and efficiently in disease prediction and other applications. However, one of the main obstacles in utilizing the medical data with big data technique is the requirement of patients' privacy protection~\cite{Shi:CnS2020}.

The adoption of blockchain to IoMT systems is a promising solution to solve the privacy problem of medical data in IoMT. The BCeMT system can preserve the privacy of patients and remove the obstacle of utilizing medical data with big data. In particular, the integration of attribute-based encryption~\cite{KZhang:TII20} and blockchains can achieve data analysis on encrypted data~\cite{WLiang:TII20}. In the future, the implementation of big data techniques in BCeMT systems can further extract useful information while preserving data privacy.

\subsection{Integration of AI with BCeMT}

Artificial intelligence, especially advances in machine learning (ML) and deep learning (DL) may potentially improve the service quality of the IoMT system. The intelligent services empowered by ML/DL may make up for the absence of medical resources. The massive IoMT data can be used to train ML/DL models so as to obtain reliable prediction models with high accuracy~\cite{ZHAO2021185}.  Once a reliable disease prediction model with high accuracy is built for the IoMT system, the spread of infectious diseases such as COVID-19 will be significantly inhibited.

Though the IoMT system continuously provides medical data,  different medical institutes need to share the encrypted medical data to obtain more data so to improve the performance of their prediction model. In this case, a large number of computational operations are required. Thus, distributed computing is a solution~\cite{hndai:IOTJM21}. In the BCeMT system, the architecture is distributed and the blockchain-based encryption operations ensure the security and privacy of medical data. In the future, task decomposition across the entire BCeMT system will be further investigated.

\section{Conclusion}
\label{sec:conclu}
In this article, we explore the application of blockchain-enabled IoMT (BCeMT) to fight against the COVID-19 pandemic. We first present a technical overview of blockchain and IoMT. We also summarize the challenges on privacy, security, and interoperability of incumbent IoMT systems. We then present the architecture of BCeMT and elaborate on the opportunities brought by BCeMT. We discuss the benefits of this architecture, including the security improvement of IoMT, privacy protection assurance of IoMT, and the interoperability amelioration of IoMT systems. We next provide several use cases of BCeMT on combating the COVID-19 pandemic. The applications of BCeMT include the prevention of infectious diseases, location sharing and contact tracing, and the supply chain of injectable medicines. Finally, we outline the future directions of BCeMT.

\bibliography{ref}
\end{sloppypar}
\end{document}